\documentclass{appolb}
\pdfoutput=1
\usepackage[usenames,dvipsnames]{color}
\definecolor{darkblue}{RGB}{0,0,196}
\usepackage[colorlinks=true,linkcolor=darkblue,citecolor=darkblue,urlcolor=darkblue]{hyperref}
\usepackage{graphicx}
\usepackage{makecell}
\usepackage{bm}
\usepackage{amsmath}

\begin{document}
\title{Dilepton production from the quark-gluon plasma using
leading-order (3+1)D anisotropic hydrodynamics
\thanks{Talk presented by Radoslaw Ryblewski at Excited QCD 2015, 8-14 March 2015, Tatranska Lomnica, Slovakia.}
}
\author{Radoslaw Ryblewski
\address{The H. Niewodnicza\'nski Institute of Nuclear Physics, Polish Academy of Sciences, PL-31342 Krak\'ow, Poland \\~}
\\ 
Michael Strickland
\address{Department of Physics, Kent State University, Kent, OH 44242 United States}
}
\maketitle
\begin{abstract}
Dilepton production from the quark-gluon plasma (QGP) phase of ultra-relativistic heavy-ion collisions is computed using the leading-order (3+1)-dimensional anisotropic hydrodynamics. It is shown that high-energy dilepton spectrum is sensitive to the initial local-rest-frame momentum-space anisotropy of the QGP. Our findings suggest that it may be possible to constrain the early-time momentum-space anisotropy in relativistic heavy-ion collisions using high-energy dilepton yields.
\end{abstract}
\PACS{11.15Bt, 04.25.Nx, 11.10Wx, 12.38Mh}
 %
\section{Introduction}
\label{sec:intro}
%
\par The enormous amount of data collected in ultra-relativistic heavy-ion collision experiments at the RHIC (Relativistic Heavy-Ion Collider) and LHC (Large Hadron Collider) have indicated that a new state of nuclear matter called quark-gluon plasma (QGP) is created in these events.  The data indicated that the QGP behaves like an almost perfect fluid, which may be, to a great extent, described within the framework of relativistic dissipative fluid dynamics (see the recent review \cite{Florkowski:2014yza}). Recently, the study of the mechanisms leading to, and the level of momentum-space isotropization/thermalization of partons comprising the QGP became an active research area \cite{Ryblewski:2013jsa,Strickland:2014pga}. Understanding of these phenomena is crucial not only for setting up the initial conditions for fluid dynamical frameworks, but also for judging applicability of the fluid dynamical approaches to the early-time dynamics of high-energy heavy-ion collisions. Various theoretical approaches have attempted to explain the mechanisms driving the QGP towards the local equilibrium state in both the strong and weak coupling limits, including gauge-gravity (AdS/CFT) duality, perturbative quantum chromodynamics (pQCD), and saturated initial-state color glass condensate (CGC) models \cite{Strickland:2014pga}. These results suggest that during the early stages of the evolution the matter possesses substantial local momentum-space anisotropies, with the transverse pressure significantly exceeding the longitudinal pressure.  In order to test these findings, one would like to find experimental observables that are sensitive to the early-time degree of momentum-space (an-)isotropy of the QGP.
\par High-energy ($E > 2$ GeV) electromagnetic probes, in particular dileptons (produced from the decay of virtual fotons) and real photons, seem to be perfect observables for this purpose, since they are produced mainly in the early stages/central region of the collision and they are weakly coupled to the QGP.  This means that they leave the system almost undisturbed (for recent reviews see \cite{Rapp:1999zw,Sakaguchi:2014ewa}). In order to study the effects of local-rest-frame (LRF) anisotropies in the system, the calculation of these observables has to be convoluted with the fluid dynamical framework which treats these effects in a reliable manner. Along these lines, in this work, we review recent results obtained by folding the dilepton production from the momentum-space anisotropic QGP with the QGP background evolution obtained using the recently developed framework of (3+1)D leading-order spheroidal anisotropic hydrodynamics \cite{
Martinez:2010sc,Florkowski:2010cf,Ryblewski:2010bs,Martinez:2010sd,
Ryblewski:2011aq,Florkowski:2011jg,Martinez:2012tu,Ryblewski:2012rr,
Florkowski:2013lza}. As reported originally in Ref.~\cite{Ryblewski:2015hea}, we find that the high-energy spectrum of dileptons is quite sensitive to the early-time momentum-space anisotropy of the QGP, and, therefore, this observable may potentially be used to experimentally constrain the degree of early-time momentum-space anisotropy.
%
\section{Dilepton rate in an anisotropic plasma}
\par At leading order in the electromagnetic coupling, ${\cal O} (\alpha^2)$, the dilepton emission rate comes from quark-antiquark annihilation.  The calculation of the rate in an anisotropic QGP was performed using relativistic kinetic theory in Refs.~\cite{Mauricio:2007vz,Martinez:2008di}. Therein, it was shown that the production rate is given by 
\begin{eqnarray}
\frac{d R^{l^+l^-}}{d^4\!P}\!\! &=&\!
\frac{5\alpha^2}{18\pi^5}\int_{-1}^1 \!d(\cos\theta_{p_1})
\!\int_{a_+}^{a_-}\!\! \frac{p_1 dp_1}{\sqrt{\chi}}\, f_q\!\left({p_1\sqrt{\!1+\!\xi\cos^2\theta_{p_1}}},\Lambda\right)
\nonumber \\
&& \hspace{0.15cm} \times f_{\bar{q}}\!\left(\sqrt{{(E\!-\!p_1)^2+\xi(p_1\cos\theta_{p_1}\!\!-P\cos\theta_P)^2}},\Lambda\right),
\label{kineticratefinal}
\end{eqnarray}
where 
\begin{eqnarray}
a_{\pm}&\equiv&\frac{M^2}{2(E-P\cos (\theta_P\pm\theta_{p_1}))} \, ,
\end{eqnarray}
and
\begin{eqnarray}
\chi \equiv (2 p_1 P \sin\theta_P\sin\theta_{p_1})^2 -[2p_1(E-P\cos\theta_P\cos\theta_{p_1})-M^2]^2 \, .
\end{eqnarray}
Above it was assumed that the three-momenta of the quark and the lepton pair are parametrized using spherical coordinates in the following way
\begin{eqnarray}
{\bf p}_1 &=&
p_1(\sin\theta_{p_1}\cos\phi_{p_1},\sin\theta_{p_1}\sin\phi_{p_1},\cos\theta_{p_1}) \, ,\nonumber\\
{\bf P} &=&
P(\sin\theta_{P}\cos\phi_{P},\sin\theta_{P}\sin\phi_{P},\cos\theta_{P}) \, ,
\end{eqnarray}
where the z-axis is oriented along the direction of anisotropy $\bf \hat{n}$ (which is taken to point along the beam line direction), and  $E$ and $M$ are the energy and the invariant mass of the lepton pair, respectively. The phase-space distribution functions of the quarks and antiquarks, which have to be specified in (\ref{kineticratefinal}) are assumed to be equal and, at leading order, given by the spheroidal Romatschke-Strickland (RS) ansatz \cite{Romatschke:2003ms}
\begin{equation}
f_{q ({\bar q})}({\bf p},\xi,\Lambda)\equiv f^{\rm iso}_{q ({\bar 
q})}(\sqrt{{\bf p}^2+\xi({\bf p\cdot \hat{n}})^2},\Lambda) \,.
\label{distansatz}
\end{equation}
In Eq.~(\ref{distansatz}) the anisotropy parameter $\xi$ describes the strength and type of anisotropy. The physical situations suggested by microscopic models discussed in Sec.~\ref{sec:intro} correspond to $\xi \geq 0$ (oblate distribution), however, we allow $\xi$ to be anywhere in the range $-1 < \xi < \infty$.  The parameter $\Lambda$ is the transverse momentum scale, which coincides with temperature in the case of local isotropic equilibrium ($\xi=0$). The isotropic distribution $f^{\rm iso}_{q ({\bar q})}$ is assumed to be a Fermi-Dirac distribution. In general, the transverse momentum scale and the anisotropy parameter may be  arbitrary functions of space-time point, which in Milne parametrization reads $X^{\mu} = (\tau \cosh \varsigma, {\bf x}_\perp, \tau \sinh \varsigma)$. Herein, we use standard notation for longitudinal proper time $\tau\equiv\sqrt{t^2 - z^2}$ and the space-time rapidity $\varsigma \equiv \tanh^{-1} (z/t)$. In this study, the  functions $\xi(X)$ and $\Lambda(X)$ are taken from from fluid dynamical simulations performed using the anisotropic hydrodynamics framework, see Sec.~\ref{sec:hydro}.
\section{Dilepton spectra}
\par With the dilepton emission rate in hand, the final invariant mass and transverse momentum spectra may be obtained by performing the following phase-space integrations 
\begin{subequations}
\begin{align}
\frac{dN^{l^+l^-}}{MdMdy}&=\int_{p_\perp^{\rm min}}^{p_\perp^{\rm max}}  p_\perp dp_\perp
   \int_{0}^{2\pi} \!\! d\phi_p
   \int \!\! d^4\!X \frac{dR^{l^+l^-}}{d^4\!P} \left(\vphantom{\frac{}{}}\xi (X), \Lambda(X)\right) , \label{Mspectrum}\\
\frac{dN^{l^+l^-}}{p_\perp dp_\perp dy}&=\int_{M^{\rm min}}^{M^{\rm max}}  M dM
   \int_{0}^{2\pi} \!\! d\phi_p 
   \int \!\! d^4\!X \frac{dR^{l^+l^-}}{d^4\!P}\left(\vphantom{\frac{}{}}\xi (X), \Lambda(X)\right)  ,\label{pTspectrum}
\end{align}
\label{spectrumeqs}
\end{subequations}
\hspace{-1mm}respectively. The four-momentum of the dilepton pair is parametrized in the following way
\begin{equation}
P^{\mu}=(m_{\perp} \cosh y, p_{\perp} \cos \phi_p, p_{\perp} \sin \phi_p, m_{\perp} \sinh y) \,.
\label{mompar}
\end{equation}
Since we work in the Milne coordinates the space-time measure is $d^4\!X=\tau d\tau \,d\varsigma \, d^2 x_\perp$. Before evaluating formulas (\ref{Mspectrum}) and (\ref{pTspectrum}) one has to remember to transform the LAB momentum of the dilepton pair to LRF (where Eq.~(\ref{kineticratefinal}) was obtained) using $p^{\prime \mu} = \Lambda^{\mu\,\,}_{\,\,\nu}\,\, p^{\nu}$. In this case, the general Lorentz boost tensor depends on the four-velocity of the local rest frame $\Lambda^{\mu\,\,}_{\,\,\nu}(u^{\rho}(X))$.
\section{Hydrodynamic evolution}
\label{sec:hydro}
\par Although the system studied here is strongly anisotropic, it is assumed that it behaves collectively, and may be effectively described using a finite set of macroscopic degrees of freedom using a fluid dynamical framework. For this purpose, we use the anisotropic hydrodynamics approach. 
\par At leading order, the evolution equations of anisotropic hydrodynamics may be derived by taking moments of the relativistic Boltzmann kinetic equation in the relaxation time approximation and using the RS ansatz (\ref{distansatz}) for the distribution function. In this way, the resulting energy-momentum conservation equations $\partial_\mu T^{\mu \nu} = 0$ and particle production equation $\partial_\mu N^{\mu} = u_\mu \left(N^{\mu}_{\rm eq}-N^{\mu}\right)/\tau_{\rm eq}$ lead to five partial differential equations for five functions of space-time, i.e, anisotropy parameter $\xi$, transverse momentum scale $\Lambda$, and three independent components of four-velocity $u^{\mu}$. The relaxation time is given through the relation $\tau_{\rm eq} = 5 \bar{\eta}/(2 T)$, where $\bar{\eta} = \eta/s$, with $\eta$ and $s$ being the shear viscosity and entropy density, respectively. The initial condition for hydrodynamic evolution of $\Lambda$ is specified on a constant proper time surface using a mixed optical Glauber model. We also assume that the initial flow in the transverse direction is vanishing, while in the longitudinal direction one initially has Bjorken flow, $v_z=z/t$. For details of the hydrodynamical calculations and values of the parameters used, we refer the reader to Sec.~4 of Ref.~\cite{Ryblewski:2015hea}.
\section{Results}
\begin{figure}[htb]
\centerline{%
\includegraphics[width=7cm]{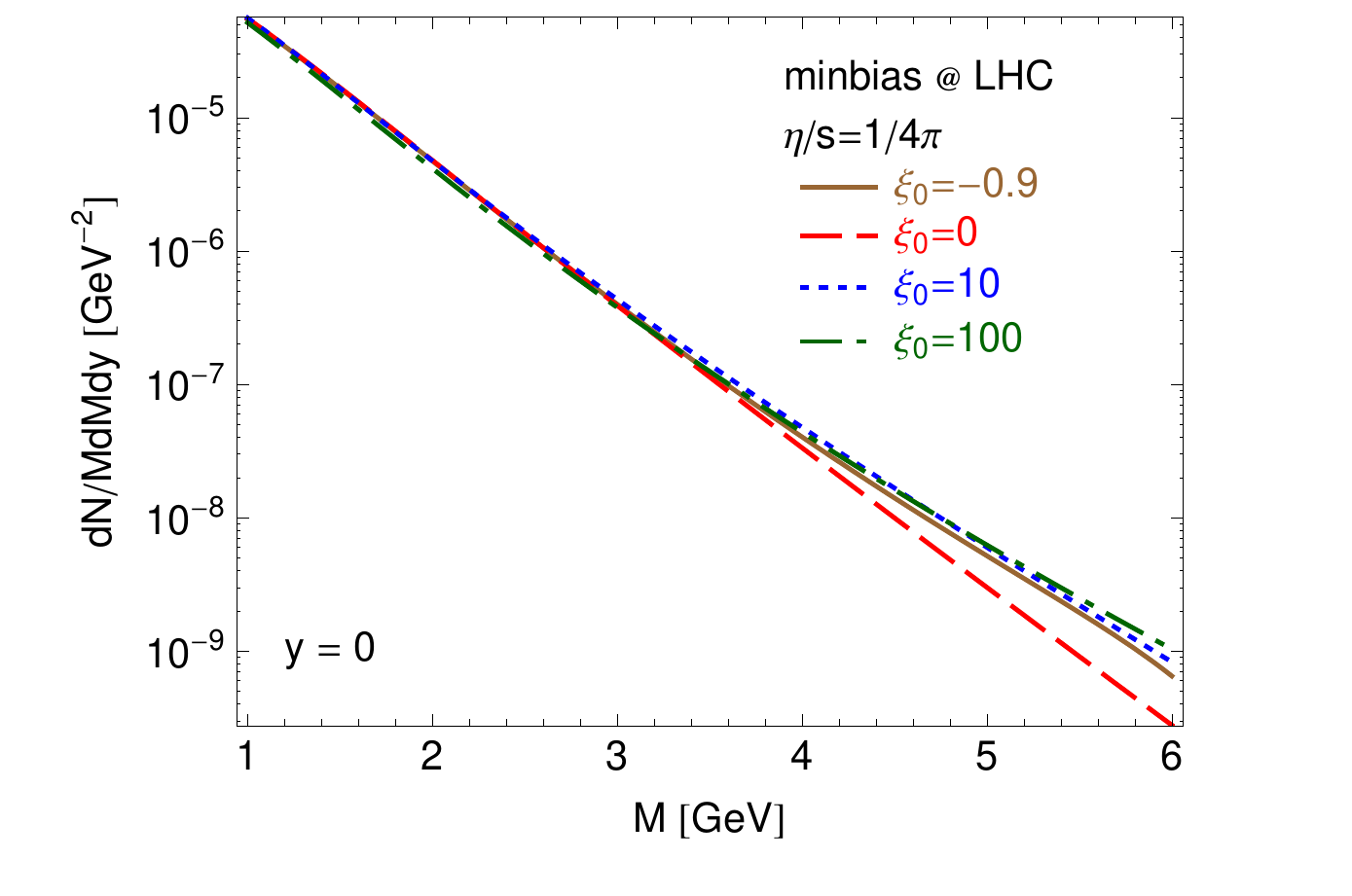}
\includegraphics[width=7cm]{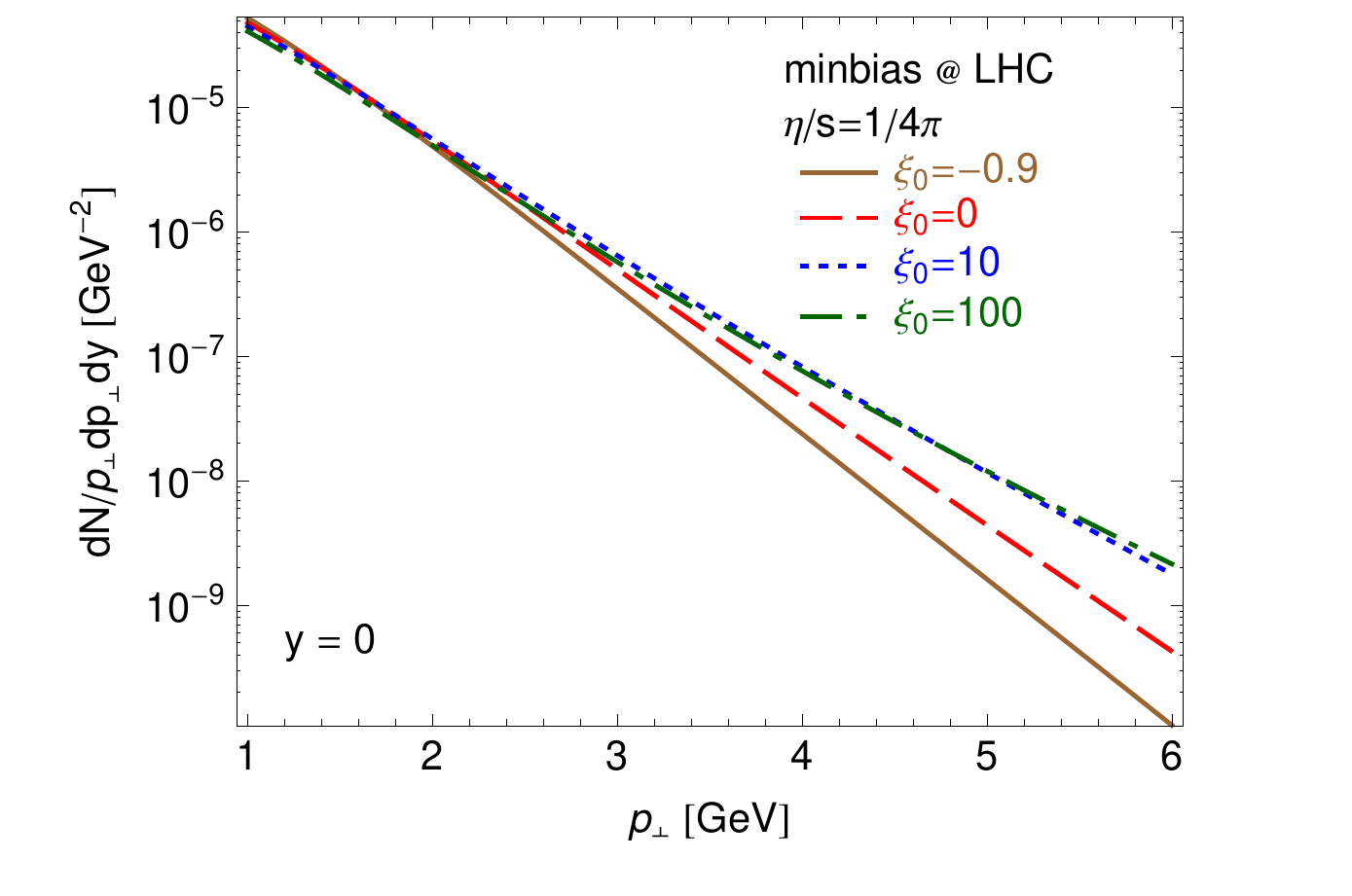}}
\caption{(Color online) The minimum-bias invariant mass (left) and transverse momentum (right) spectra of dilepton pairs calculated within anisotropic hydrodynamics for various values of the initial anisotropy $\xi_0$ and the AdS/CFT lower bound value of shear viscosity to entropy density ratio, $\eta/s = 1/(4\pi)$.}
\label{fig:spectra}
\end{figure}

The hydrodynamic model presented in previous Section is used to simulate a (3+1)-dimensional minimum-bias lead-lead collision at LHC beam energy of $2.76$ GeV. After the hydrodynamic evolution is determined the dilepton spectra are computed using formulas (\ref{Mspectrum}) and (\ref{pTspectrum}), assuming that the emission stops when the temperature drops below a critical temperature of $175$ MeV. In the calculations, we restrict ourselves to the high-energy part of the spectrum by setting the cuts to $p_\perp^{\rm min} =1$ GeV and $p_\perp^{\rm max} = 20$ GeV for transverse momentum integration, and $M^{\rm min} = 1$ GeV and $ M^{\rm max} = 20$ GeV for the invariant mass integration. In Fig.~\ref{fig:spectra}, we present exemplar results for the invariant mass (left) and transverse momentum (right) spectra for various values of the initial momentum-space anisotropy. One observes that the transverse momentum spectra in particular is extremely sensitive to the initial anisotropy assumed. We observe that, with increasing initial anisotropy, the spectra becomes flatter. We checked that the observed effects are larger for $\eta/s > 1/4\pi$.
%
\section{Conclusions}
Based on the model results obtained, we find that the invariant mass and transverse momentum spectra are sensitive to the initial anisotropy of the QGP. Our findings provide possibility to constrain the level of isotropisation of the strongly interacting medium created in relativistic heavy-ion collisions by measuring the high-energy part of the dilepton spectra.  It would be very interesting to confront our findings with the experimental data including additional dilepton emission from the hadronic phase. Moreover, it would be interesting to go beyond the leading-order formulation of the anisotropic hydrodynamics by including NLO corrections \cite{Bazow:2013ifa,Bazow:2015cha}, or to use more complete leading-order formulations of anisotropic hydrodynamics \cite{Tinti:2013vba,Tinti:2014yya}.

\section*{Acknowledgments}
R.R. would like to acknowledge the organizers of Excited QCD 2015 for their hospitality. R.R. was supported by Polish National Science Center Grant No. DEC-2012/07/D/ST2/02125.  M.S. was supported by U.S. DOE Award No. DE-AC0205CH11231.

\end{document}